# Regression Models for Order-of-Addition Experiments

Hans-Peter Piepho[a,*], Emlyn R. Williams[b]

[a] *Biometrics Unit, University of Hohenheim, Stuttgart, Germany*

[b] *Statistical Consulting Unit, Australian National University, Canberra, Australia*

CONTACT: Hans-Peter Piepho (piepho@uni-hohenheim.de) Institute of Crop Science, Biometrics Unit, University of Hohenheim, 70593 Stuttgart, Germany

# Regression Models for Order-of-Addition Experiments

The purpose of order-of-addition (OofA) experiments is to identify the best order in a sequence of *m* components in a system or treatment. Such experiments may be analysed by various regression models, the most popular ones being based on pairwise ordering (PWO) factors or on component-position (CP) factors. This paper reviews these models and extensions and proposes a new class of models based on response surface (RS) regression using component position numbers as predictor variables. Using two published examples, it is shown that RS models can be quite competitive. In case of model uncertainty, we advocate the use of model averaging for analysis. The averaging idea leads naturally to a design approach based on a compound optimality criterion assigning weights to each candidate model.

## 1. Introduction

The purpose of order-of-addition (OofA) experiments is to identify the best order in a sequence of *m* components in a system or treatment (Van Nostrand 1995). For example, *m* herbicides may need to be combined to obtain a herbicide mixture targeting a range of weed species and it is not clear in which order the mixture components should be added to the tank (Mee 2020). Another example is a medical treatment involving *m* drugs, and the optimal order of administration needs to be determined (Table 1; Yang et al. 2020).

[Table 1 about here]

With *m* components, there are *m*! possible orders (see Table 1). The full design comprises one run for each possible order. When *m* is large, only a subset of the possible orders may be tested. With smaller *m*, replication of some or all orders may be feasible, thus providing an independent estimate of error. The choice of design for OofA experiments has received considerable attention recently. Most authors focus on D-optimal designs, primarily based on combinatorics for design generation, e.g. using block designs (Chen et al. 2020) or component orthogonal arrays (Yang et al. 2020; Zhao et al. 2020; Huang 2021), whereas other authors proposed purely numerical search strategies (Voelkel, 2019; Mee 2020, Winkler and Lin 2020).

There are currently two popular basic models for analysis, i.e., a model using pairwise ordering (PWO) factors (Van Nostrand 1995) and the component-position (CP) model (Yang et al. 2020). The PWO model has linear predictor

$$\eta = \beta_0 + \sum_{c=1}^{m-1} \sum_{d=c+1}^{m} \beta_{cd} x_{cd} \qquad (1)$$

where for each pair $(c,d)$ of components $c$ and $d$ $(c<d)$, $x_{cd} = 1$ if component $c$ precedes component $d$ and $x_{cd} = -1$ when component $d$ precedes component $c$. This model has $1 + m(m-1)/2$ parameters and focuses on the relative position of components to one another. Extensions of this model were considered, e.g., by Peng et al. (2019), Lin and Peng (2019), and Mee (2020). By contrast, the CP model focuses on the absolute positions. It is given by

$$\eta = \beta_0 + \sum_{c=1}^{m} \sum_{j=1}^{m} x_c^{(j)} \tau_c^{(j)} \qquad (2)$$

where $x_c^{(j)} = 1$ if component $c$ is used at position $j$ and 0 otherwise, and $\tau_c^{(j)}$ is the effect of the $c$-th component at position $j$. The model has $1 + (m-1)^2$ free parameters, observing that there are $2m-1$ constraints on the parameters $\tau_c^{(j)}$. For example, we may set $\tau_c^{(j)} = 0$ for $j = m$ or $c = m$ (baseline constraints).

Both the PWO and the CP model imply that the position of a component in the sequence matters, but they differ in how position is thought to affect the outcome. The PWO focuses on the position of components relative to each other, whereas the CP model focuses on the absolute positions of each component in the sequence. Which model is preferable will depend on the application. We conjecture, however, that for either model it will be beneficial to have components well spread out across positions over the runs of a design so that the optimal position of each component can be determined.

The assumptions underlying the PWO and CP models are certainly plausible in principle, but they may not always be the best possible options. For use of OofA experiments in practice, it is useful to have a number of alternative models for analysis (Buckland et al. 1997). The purpose of this paper is to consider such alternative models and illustrate them using examples. The focus will be on regression models that use the position of components to define regressor variables. In particular, we will propose response-surface (RS) regression models, which can be thought of as accounting for both relative and absolute positions of the components at the same time. The general picture emerging from our review is that there is potentially quite a substantial number of regression models. Hence, we advocate model averaging (Buckland et al. 1997) as a viable analysis option. In the discussion, we will also consider the implications for design.

**2. Regression models**

*2.1 A second-order response surface model*

The CP model assumes that the effect of a component $c$ depends solely on its absolute position $q_c \in (1,2,...,m)$ in the sequence. If this premise is accepted, it may be postulated that the effect varies according to a smooth function of position number $q_c$ according to a response-surface (RS) model. This suggests that some nonlinear regression model may be employed that uses position numbers for all components as regressors. In particular, it may be postulated that there is an optimal position of each component on the position scale. There are of course many potential regression models for this purpose. Here, we will focus on polynomials for their simplicity and flexibility (Box and Draper 2007).

For each run, the position numbers add up to $\sum_{c=1}^{m} q_c = \sum_{j=1}^{m} j = m(m+1)/2$. It will be convenient to standardize position numbers so they add up to unity for each run, i.e.,

$$p_c = \frac{2q_c}{m(m+1)} \qquad (3)$$

so that for every run

$$\sum_{c=1}^{m} p_c = 1. \qquad (4)$$

It transpires from (4) that the regression problem is akin to that of finding an optimal mixture of $m$ components, with $p_c$ taking on the role of mixing proportions (Box and Draper 2007, p.509). With OofA experiments, however, there are further constraints on the permissible values for $p_c$, and this has implications for the regression model. Specifically, we have an additional constraint on the quadratic terms:

$$\sum_{c=1}^{m} p_c^2 = \frac{2(2m+1)}{3m(m+1)} \;. \qquad (5)$$

Similarly, it may be shown for the cross-product terms that

$$\sum_{c=1}^{m-1} \sum_{d=c+1}^{m} p_c p_d = \frac{3m^2 - m - 2}{6m(m+1)} \;. \qquad (6)$$

We may initially assume a second-order RS model of the form (Box and Draper 2007)

$$\eta = \beta_0 + \sum_{c=1}^{m} \beta_c p_c + \sum_{c=1}^{m} \beta_{cc} p_c^2 + \sum_{c=1}^{m-1} \sum_{d=c+1}^{m} \beta_{cd} p_c p_d \qquad (7)$$

This model can be reduced in three steps to account for the constraints on $p_c$, in a similar fashion as for mixture experiments (Box and Draper 2007, p.513 and p.517).

First, we replace the last component's position in the linear term with $p_m = 1 - \sum_{c=1}^{m-1} p_c$ and note that the resulting term $1 \times \beta_m$ can be absorbed into the intercept $\beta_0$. Likewise, the last component's quadratic position term is replaced with $p_m^2 = \frac{2(2m+1)}{3m(m+1)} - \sum_{c=1}^{m-1} p_c^2$, where again the resulting constant term $\frac{2(2m+1)}{3m(m+1)} \times \beta_{mm}$ is absorbed into the intercept.

Thus, the reduced model is

$$\eta = \beta_0 + \sum_{c=1}^{m-1} \beta_c p_c + \sum_{c=1}^{m-1} \beta_{cc} p_c^2 + \sum_{c=1}^{m-1} \sum_{d=c+1}^{m} \beta_{cd} p_c p_d \ . \tag{8}$$

Second, it follows from (6) that the intercept can be absorbed into the cross-product terms, leading to the model

$$\eta = \sum_{c=1}^{m-1} \beta_c p_c + \sum_{c=1}^{m-1} \beta_{cc} p_c^2 + \sum_{c=1}^{m-1} \sum_{d=c+1}^{m} \beta_{cd} p_c p_d \ . \tag{9}$$

Third, we replace the position of the last component by $p_m = 1 - \sum_{c=1}^{m-1} p_c$ in the cross-product terms $p_m p_d$ $(d \neq m)$. Observing that the resulting terms can be absorbed into the remaining linear, quadratic and cross-product terms, we obtain the equivalent second-order OofA model

$$\eta = \sum_{c=1}^{m-1} \beta_c p_c + \sum_{c=1}^{m-1} \beta_{cc} p_c^2 + \sum_{c=1}^{m-2} \sum_{d=c+1}^{m-1} \beta_{cd} p_c p_d \ . \tag{10}$$

In summary, this is the usual second-order response surface model, reduced by the intercept and all terms involving $p_m$. For an alternative derivation and third-order extensions, see Appendix A. This model has $(m-1)(m+2)/2$ parameters, which is halfway between the PWO and CP models in terms of complexity. For example, for $m = (3,4,5,6)$ the number of parameters are $(4,7,11,16)$, $(5,9,14,20)$ and $(5,10,17,26)$ for the PWO, RS and CP models, respectively.

It should be pointed out that a regression on absolute positions $p_c$ ($q_c$) of the components can be considered as also accounting for relative position. This is because the $m$ absolute positions $q_c$ can be re-expressed by the absolute position of one common reference and $m-1$ distances relative to the common reference. For example, taking the $m$-th component as the reference, we may replace $q_c$ $(c=1,...,m-1)$ with $q_c = q_m + \delta_{cm}$, where $\delta_{cm} = q_c - q_m$ is the signed distance of component $c$ from reference component $m$, with $\delta_{cm} < 0$ $(\delta_{cm} > 0)$ implying that component $c$ precedes (follows) component $m$. If this replacement is made in (10), we obtain a second-order model in the distances $\delta_{cm}$, meaning that the models allows for an optimal position for each $c < m$ relative to the reference component $m$.

## *2.2 Modifications of the PWO model: A nearest neighbour model and a tapered-effects model*

The PWO model implies that in judging two components $c$ and $d$, it only matters whether component $c$ comes before or after $d$, regardless of how close they are. This view can be modified in different ways. One option is to focus on immediate neighbours. This leads to a nearest-neighbour (NN) model with linear predictor

$$\eta = \sum_{c=1}^{m} \sum_{d \neq c}^{m} \beta_{cd} w_{cd} , \qquad (11)$$

where for each pair $(c,d)$ of components $c$ and $d$, $w_{cd} = 1$ if component $c$ immediately precedes component $d$ and $w_{cd} = 0$ otherwise. Note that the model has no intercept as $\sum_{c=1}^{m} \sum_{d \neq c}^{m} w_{cd} = m-1$ for any possible order. This NN model has $m(m-1)$ parameters and focuses on the relative position of components to one another, as does the PWO model.

A further option is to assume that in the PWO model the effect size decays with distance $h_{cd} = |q_c - q_d|$. Peng et al. (2019) propose two nonlinear functions for such tapered effects, i.e. $z(h) = 1/h$ and $z(h) = c^{h-1}$ for some known $c$ such that $0 < c < 1$. Another option is the linear function $z(h) = m - h$. Using in (1) any of these choices and setting $x_{cd} = z(h_{cd})$ when $c$ precedes $d$ and $x_{cd} = -z(h_{cd})$ otherwise yields what we will call the tapered PWO (*t*PWO) model. It turns out that with the linear function $z(h) = m - h$, the *t*PWO model yields the same fit as the ordinary PWO model (1) (see Appendix B). This, in turn, suggests that other choices for $z(h)$ will not provide much improvement of the *t*PWO model over the PWO model (1).

**2.3 Criteria for judging the predictive accuracy of a model for a given design**

When $m$ is large ($\geq 10$, say), even identifying the best order(s) from a fitted model may be computationally far from trivial, and the best strategy may well depend on the kind of model. In this paper, we will focus on the situation when $m$ is small ($m < 7$, say). Then an obvious strategy is to predict the response for all possible orders and then perform all pairwise comparisons between the predictions in order to find the best one. One may simply use the order with the best prediction. Alternatively, a subset of best solutions could be identified, and the choice substantiated by significance tests. Specifically, Hsu's multiple comparison with the best (Hsu 1996, p.81) could be used to account for the multiple testing problem involved in identifying this subset. This approach is completely general and would work for any regression model. But it will become infeasible even for modest $m$, e.g. $m = 10$, where $m! = 3,628,800$. Even when predictions are not actually computed for all possible orders, however, identification of the best order still comes down, at least implicitly, to comparing the predictions of all possible orders.

Thus, regardless of the value of *m*, in judging the efficiency of a design, it is appropriate to consider the predictions for all *m*! orders. Let the linear predictor for a design be given by $X\beta$, where $\beta$ is the parameter vector, and let the full set of *m*! orders be represented by the matrix $X_f$. Thus, we are aiming to predict $\eta_f = X_f \beta$, for which the least-squares estimator is $\hat{\eta}_f = X_f \hat{\beta} = X_f (X^T X)^{-1} X^T Y$, where *Y* is the observed data vector of length *N* for the design. Assuming that $\text{var}(Y) = I_N \sigma^2$, the variance of the prediction is $\text{var}(\hat{\eta}_f) = X_f (X^T X)^{-1} X_f^T \sigma^2$. From this, we may obtain the average pairwise variance of predictions as (Piepho 2019)

$$apv = \frac{2\sigma^2}{w-1} trace\left[ (X^T X)^{-1} X_f^T (I_w - w^{-1} J_w) X_f \right], \qquad (12)$$

where $w = m!$ is the number of rows in $X_f$. While $X_f$ is a huge matrix, the matrix $X_f^T (I_w - w^{-1} J_w) X_f$ is just a $p \times p$ matrix, where *p* is the number of parameters in $\beta$. When *apv* is to be used in design search, we may set $\sigma^2 = 1$ and the matrix $X_f^T (I_w - w^{-1} J_w) X_f$ needs to be computed only once. We note that the *apv* bears some resemblance with the I-optimality criterion (Goos et al. 2016), sometimes also denoted as V-optimality (Atkinson et al. 2007, p.143), which focuses on the average variance (*av*) of predictions over the experimental region. For the full set of *m*! orders in a OofA experiment this equals

$$av = \frac{\sigma^2}{w} trace\left[ (X^T X)^{-1} X_f^T X_f \right], \qquad (13)$$

whereas (12) focuses on the prediction variance of the pairwise differences across all possible pairs of orders. By comparison, an A-optimal design minimizes $\frac{\sigma^2}{p} trace\left[ (X^T X)^{-1} \right]$, whereas a D-optimal design maximizes $\sigma^2 |X^T X|^{1/p}$ (Atkinson et al.

2007, p.137; Jones et al. 2020). Neither of these latter two criteria focuses directly on the prediction of $\eta_f = X_f \beta$. It is noteworthy that the A-optimality criterion is not invariant to equivalent re-parameterizations (codings) of the linear predictor. One particularly useful re-parameterization is where the design matrix for the full set of $m!$ runs is orthogonally coded such that $X_f^T X_f = w I_w$, in which case $av = \sigma^2 trace\left[(X^T X)^{-1}\right]$, corresponding to the A-optimality criterion (SAS Institute, 2014, p.1046). Orthogonalization can be done by finding a square matrix $R$ such that $X_f^T X_f = R^T R$, e.g. by Gram-Schmidt orthgonalization, and then replacing each row $x_f$ of $X_f$ by $x_f R^{-1} \sqrt{w}$, as is done in OPTEX when the option CODING=ORTH is used (Atkinson et al. 2007, p.188).

*2.4 Model averaging*

Model choice may not be obvious, in which case model averaging (Buckland et al. 1997; Burnham and Anderson 2002, p.450) is a good option for obtaining predictions of $\eta_f = X_f \beta$. A model-averaged prediction is given by (Buckland et al. 1997)

$$\hat{\eta}_f = \sum_k w_k \hat{\eta}_{f,k}, \tag{14}$$

where $\hat{\eta}_{f,k}$ is the prediction based on the $k$-th model $g_k$ $(k = 1,...,K)$ and $w_k$ is the weight of the $k$-th model, subject to the constraint $\sum w_k = 1$. One choice of weights is given by

$$w_k = \frac{\exp(-I_k/2)}{\sum_{h=1}^{K} \exp(-I_h/2)}, \tag{15}$$

where $I$ denotes an information criterion such as the Akaike Information Criterion (AIC). The variance of the model-averaged prediction of the $i$-th order of application

can be estimated from (Burnham and Anderson 2002, p. 450)

$$\text{var}(\hat{\eta}_{f(i)}) = \left\{ \sum_k w_k \sqrt{\text{var}(\hat{\eta}_{f,k(i)} \mid g_k) + (\hat{\eta}_{f,k(i)} - \hat{\eta}_{f(i)})^2} \right\}^2 \qquad (16)$$

For alternatives to estimate both the weights and the variance of predictions, see Buckland et al. (1997). To evaluate the overall performance of the prediction, (16) may be averaged across the complete set of application orders.

*2.5 Examples*

In Figure 1, the second-order response surface model (10) is illustrated for $m = 3$ using the data of Table 1. Circles correspond to design points; circle filled with red is the component order largest predicted response.

[Figure 1 about here]

We further consider the data for $N = 24$ runs and $m = 4$ anti-tumor drugs in Table 3 of Yang et al. (2020). The fitted models are reported in Table 2. The RS model yields the best fit in terms of the root mean square of error (RMSE) and the Akaike Information Criterion (AIC) (Burnham and Anderson, 2002), whereas the PWO and *t*PWO models has a smaller *avd*. The model-averaged predictions for the ten best combinations is shown in Table 3. The RS model has the largest Akaike weight ($w_k = 0.410$), closely followed by the *t*PWO model ($w_k = 0.376$) (Table 2). In this example, it would be hard to make a choice among these two best-fitting models, so the model-averaged inference is very apt. The ranking of the model-averaged predictions for the ten best combinations agrees rather well with those obtained for the individual models, though there are some rank changes (Table 3).

[Tables 2 and 3 about here]

A third example is the data given in Table 4 of Yang et al. (2020) from an experiment with $N = 40$ runs and $m = 5$ anti-tumor drugs administered one after the other every three hours. The model fits reported in Table 4 show that the RS model fits best and rather better than the second-best CP model. The very large Akaike weight ($w_k$ = 0.99996) for the RS model means that this dominates the model-averaged prediction in this example. The ranking of both the model-averaged prediction and the prediction based on the RS model, which agree perfectly, are rather different from those based on the other models (Table 5).

[Tables 4 and 5 about here]

## 4. Discussion

### *4.1 Model selection*

Our review has revealed that there are several candidate regression models. This raises the question, which model should be used for analysis. The answer will very much depend on the application. If the best suitable model is known *a priori* for the application at hand, both the design and the analysis may proceed considering just that model. But what if model choice is not clear *a priori*? As regards analysis, one obvious option is to fit all candidate models and pick the best one based on some standard criterion such as information criteria (Burnham and Anderson 2002) or *apv*. A better strategy to deal with model-selection uncertainty is model-averaging (Buckland et al. 1997), as illustrated briefly in Section 2.5.

The set of candidate models can be expanded in many ways. For example, each of the regression models considered in our review may be extended or modified. Mee (2020) suggested extending the PWO model by including interaction terms, e.g. for triplets of components applied in sequence. The *t*PWO model of Peng et al. (2019) gives rise to a whole family of models depending on the choice of the tapering function $z(h)$. It could also be postulated that the tapering should be different for components preceding a position and those following a position, calling for a directional version of $z(h)$. Conversely, our NN model (11) assumes that direction matters for the neighbour effects. This could be simplified by assuming a non-directional NN model. We do not want to expand this list of examples further, but merely re-iterate that there are potentially many candidate models.

A further option to expand the candidate set substantially is variable selection, e.g. by stepwise regression. This has been considered by various authors, e.g., Mee (2020) and Yang et al. (2020). With some proposed models, such as when higher-order interaction terms are included (Mee 2020), variable selection is a necessity when the number of runs is limited. Again, rather than settling for a single final model, which is a decision that is always subject to uncertainty and entails the risk of over- or underfitting (Burnham and Anderson, 2002; Heinze et al. 2018), model averaging may be a better strategy. It also seems prudent to limit the candidate set by making best use of prior knowledge and focus on the more parsimonious models.

### *4.2 Implications for design generation*

There is a growing body of literature on optimal design for OofA experiments, much of which focuses on a single model, which is the obvious strategy when the best design is known with certainty *a priori*. For example, Winkler et al. (2020) and Chen et al. (2020) focus on the PWO model, while Huang (2021) focuses on the CP model. Yang et al. (2020) and Zhao et al. (2020) consider both of these models in finding a design, first constructing component orthogonal arrays. These designs are globally optimal under the CP model. Then among these designs, they select the best designs in terms of D-optimality for the PWO model. If there is uncertainty as regards the best model for analysis and there is a larger set of candidate models, however, it may not be sensible to optimize the design for particular models. Instead, a compound criterion could be optimized that averages an optimality criterion across a select set of candidate models. This idea has an obvious counter-part in analysis, i.e., model averaging (Buckland et al. 1997) (see sections 2.4 and 2.5).

Thus, for design generation, it seems reasonable to consider two quite different scenarios: (i) The most suitable model is known *a priori*; (ii) the best model is not known and a set of candidate models will be considered for analysis. In scenario (i), the obvious design strategy is to find an optimal design, e.g. one that minimizes *apv*, for the known best model. In scenario (ii), a compound criterion covering the set of candidate models may be used to find an optimal design. If $M_k(\xi)$ and $\Psi_k\{M_k(\xi)\}$ denote the information matrix and optimality criterion for the *k*-th model $(k=1,...,K)$, respectively, then the compound optimality criterion of the design $\xi$ is

$$\Psi(\xi) = \sum_{k=1}^{K} a_k \Psi_k\{M_k(\xi)\}, \qquad (14)$$

where $a_k$, $k=1,...,K$ is a set of non-negative weights (Atkinson et al., 2007, p.144). Without any prior knowledge, all models in the candidate set may receive equal weight $a_k = 1/k$. Alternatively, weights may differ depending on prior preference, e.g. based on previous estimates of Akaike weights $w_k$ (see sections 2.4 and 2.5). An additional challenge with the search under scenario (ii) is that it needs to be made sure that each model in the candidate set is estimable for a selected design (Smucker et al. 2012).

We think that with OofA experiments the best choice for $\Psi_k\{M_k(\xi)\}$ is *apv*. This may be computationally somewhat demanding because the information matrix $M_k(\xi)$ [corresponding to $(X^T X)^{-1}$ in (12)] needs to be updated in each step of a design search. A computationally cheaper option is D-optimality because efficient update equations exist for the determinant of $M_k(\xi)$ (Dykstra 1971). D-optimal designs also usually do quite well for other optimality criteria including I-optimality (Atkinson et al. (2007, p.153). Donev and Atkinson (1988) showed this for response surface designs.

**Appendix A: An alternative derivation of the second-order response surface model and third-order extensions**

For mixtures, observing the linear constraint (4), the second-order RS model can be written in canonical form as (Scheffé 1958, 1963)

$$\sum_{c=1}^{m} \beta_c p_c + \sum_{c=1}^{m-1} \sum_{d=c+1}^{m} \beta_{cd} p_c p_d \tag{A1}$$

This model has $m(m+1)/2$ parameters. For OofA experiments, we need to observe the additional constraint (6), which means that, e.g., the last cross-product of positions can be replaced by

$$p_{m-1} p_m = \frac{3m^2 - m - 2}{6m(m+1)} - \sum_{c=1}^{m-2} \sum_{d=c+1}^{m} p_c p_d \tag{A2}$$

The cross-product terms on the right-hand side of (A2) will be absorbed into the like terms in (A1), whereas the constant generates an intercept term, which in turn can be absorbed into the linear terms of (A1). Thus, the second-order RS model is

$$\sum_{c=1}^{m} \beta_c p_c + \sum_{c=1}^{m-2} \sum_{d=c+1}^{m} \beta_{cd} p_c p_d \tag{A3}$$

This model has $m(m+1)/2 - 1 = (m-1)(m+2)/2$ parameters, as does model (10), and in fact both are equivalent.

More complex models than the second-order model may be needed in some cases. The canonical second-order model (10) for mixtures has a third-order canonical extension given by Box and Draper (2007, p.520) as

$$\sum_{c=1}^{m}\beta_c p_c + \sum_{c=1}^{m-1}\sum_{d=c+1}^{m}\beta_{cd} p_c p_d + \sum_{c=1}^{m-1}\sum_{d=c+1}^{m}\alpha_{cd} p_c p_d (p_c - p_d) + \sum_{c=1}^{m-2}\sum_{d=c+1}^{m-1}\sum_{e=d+1}^{m}\beta_{cde} p_c p_d p_e \quad (A4)$$

This model has $m(m^2-1)/6$ more terms than model (10). Observing the additional constraints arising from the fact $\sum_{c=1}^{m-2}\sum_{d=c+1}^{m-1}\sum_{e=d+1}^{m} p_c p_d p_e$ is constant (note that the value of the constant is immaterial for the argument), the terms $\beta_{m-1,m}$ and $\beta_{m-2,m-1,m}$ can be dropped. The other third-order term, $\alpha_{cd} p_c p_d (p_c - p_d)$, requires dropping $(m-1)$ terms. For example, we may collect all terms involving $p_m$, yielding

$$p_m \sum_{c=1}^{m-1} p_c^2 - p_m^2 \sum_{c=1}^{m-1} p_c = p_m \frac{2(2m+1)}{3m(m+1)} - p_m\left(1 - \sum_{c=1}^{m-1} p_c\right) . \quad (A5)$$

All terms on the right-hand side are absorbed into the linear and bilinear terms ($\beta_c$ and $\beta_{cd}$) and so can be dropped. Hence, the resulting third-order RS model is

$$\sum_{c=1}^{m}\beta_c p_c + \sum_{c=1}^{m-2}\sum_{d=c+1}^{m}\beta_{cd} p_c p_d + \sum_{c=1}^{m-2}\sum_{d=c+1}^{m-1}\alpha_{cd} p_c p_d (p_c - p_d) + \sum_{c=1}^{m-3}\sum_{d=c+1}^{m-1}\sum_{e=d+1}^{m}\beta_{cde} p_c p_d p_e . \quad (A6)$$

This model has $m(m^2-1)/6 - m$ model terms than the second-order OofA model (A3).

There is also a reduced version of the full cubic model (A4) for mixtures, known as the special cubic model (Box and Draper 2007, p.520), obtained by dropping all of the $\alpha_{cd}$ terms, which has a corresponding special cubic counterpart to the RS model in (A6), when all the $\alpha_{cd}$ terms are dropped:

$$\sum_{c=1}^{m}\beta_c p_c + \sum_{c=1}^{m-2}\sum_{d=c+1}^{m}\beta_{cd} p_c p_d + \sum_{c=1}^{m-3}\sum_{d=c+1}^{m-1}\sum_{e=d+1}^{m}\beta_{cde} p_c p_d p_e . \quad (A7)$$

The second-order model and these third-order models are nested and so may be compared by F-tests for nested models. It does not seem that these models provide an improvement compared to the fit of the second-order RS model for the two examples in Tables 2 and 4 (see Tables A1 and A2).

[Tables A1 and A2 about here]

**Appendix B: One-to-one mapping between PWO model and linear *t*PWO model**

*Theorem*: The PWO model and the linear *t*PWO model yield identical predictions.

*Proof*: The PWO model is parameterized in terms of the covariate $x_{cd}$ for $c < d$. For convenience, we here consider $x_{cd}$ for all $c \neq d$, observing that $x_{cd} = -x_{dc}$. Also, we will consider the signed distance $\tilde{h}_{cd} = q_d - q_c$ with $\tilde{h}_{cd} > 0$ when component *c* precedes component *d*, i.e., $\tilde{h}_{cd} = x_{cd} h_{cd}$. It is readily verified that

$$\tilde{h}_{cd} = \frac{1}{2}\left(\sum_{e \neq c} x_{ce} - \sum_{e \neq d} x_{de}\right). \tag{B1}$$

The covariate for the linear tapered PWO can be computed from those for the PWO model as

$$x_{cd} z(h_{cd}) = x_{cd}(m - h_{cd}) = mx_{cd} - \tilde{h}_{cd} = (m-1)x_{cd} - \frac{1}{2}\left(\sum_{e \neq c,d} x_{ce} - \sum_{e \neq c,d} x_{de}\right), \tag{B2}$$

which is obviously a linear mapping $x_{cd} \to x_{cd} z(h_{cd})$. Next we show that this mapping is one-to-one.

Denote the design matrixes of the PWO and the linear *t*PWO model as $X_{PWO}$ and $X_{ltPWO}$, respectively, and denote their number of columns by *p*. The linear mapping (B2) can be written as $X_{ltPWO} = X_{PWO} A$ for a suitable matrix $A$. The inverse mapping is $X_{PWO} = X_{ltPWO} B$. We have from standard results on the rank of matrices (Harville, 1997, p. 396) that $rank(X_{ltPWO} = X_{PWO} A) \leq \min[rank(X_{PWO}), rank(A)]$. Since $rank(X_{ltPWO}) = rank(X_{PWO}) = p$, it follows that $rank(A) = p$ and likewise that $rank(B) = p$. Hence the linear mapping is one-to-one.

*Corollary*: $A = \left(X_{PWO}^T X_{PWO}\right)^{-1} X_{PWO}^T X_{ltPWO}$ and $B = \left(X_{ltPWO}^T X_{ltPWO}\right)^{-1} X_{ltPWO}^T X_{PWO}$.

**References**

Atkinson, A. C., Donev, A. N., and Tobias, R. D. (2007), *Optimum Experimental Designs, with SAS*, Oxford: Oxford University Press.

Box, G. E. P., and Draper, N. R. (2007), *Response Surfaces, Mixtures, and Ridge Analysis*, New York, NY: Wiley.

Buckland, S. T., Burnham, K. P., and Augustin, N. H. (1997), "Model Selection: An Integral Part of Inference," *Biometrics*, 53, 603−618.

Burnham, K. P., and Anderson, D. R. (2002), *Model Selection and Multimodel Inference. A Practical Information-Theoretic Approach, Second edition*. New York: Wiley.

Chen, J., Mukerjee, R., and Lin, D. K. (2020), "Construction of Optimal Fractional Order-of-Addition Designs via Block Designs," *Statistics & Probability Letters*, 161, 108728.

Cook, R. D., and Nachtsheim, C. J. (1980), "A Comparison of Algorithms for Constructing Exact D-Optimal Designs," *Technometrics*, 22, 315–324.

Donev, A. N., and Atkinson, A. C. (1988), "An Adjustment Algorithm for the Construction of Exact D-optimum Experimental Designs," *Technometrics*, 30, 429–433.

Dykstra, O., Jr. (1971), "The Augmentation of Experimental Data to Maximize $|X'X|$," *Technometrics*, 13, 682–688.

Goos, P., Jones, B., and Syafitri, U. (2016), "I-optimal Design of Mixture Experiments," *Journal of the American Statistical Association*, 111, 899–911.

Harville, D. A. (1997), *Matrix Algebra for Statisticians*, New York: Springer.

Heinze, G, Wallisch, C., and Dunkler, D. (2018), "Variable Selection – A Review and Recommendations for the Practicing Statistician. Biometrical Journal 60, 431-449

Hsu, J. C. (1996), *Multiple Comparisons. Theory and Methods*, London: Chapman and Hall.

Huang, H. (2021), "Construction of Component Orthogonal arrays with Any Number of Components," *Journal of Statistical Planning and Inference*, 213, 72–79.

Jin, D. K. J., and Peng, J. (2019), "Order-of-Addition Experiments: A Review and Some New Thoughts," *Quality Engineering*, 31, 49–59.

Jones, B., Allen-Moyer, K., and Goos, P. (2020), "A-Optimal Versus D-Optimal Design of Screening Experiments," *Journal of Quality Technology*, , –.

Mee, R. W. (2020), "Order-of-Addition Modelling," *Statistica Sinica*, 30, 1543–1559.

Peng, J., Mukerjee, R., and Lin, D. K. J. (2019), "Design of Order-of-Addition Experiments," *Biometrika*, 106, 683–694.

Piepho, H.P. (2019), "A Coefficient of Determination ($R^2$) for Generalized Linear Mixed Models," *Biometrical Journal*, 61, 860–872.

SAS Institute Inc. (2014), *SAS/QC®13.2 User's Guide*, Cary, NC: SAS Institute Inc. https://support.sas.com/documentation/onlinedoc/qc/132/optex.pdf

Scheffé, H. (1958), "Experiments with Mixtures," *Journal of the Royal Statistical Society B*, 20, 344–360 (correction: 1959: 238).

Scheffé, H. (1963), "The Simplex-Centroid Design for Experiments with Mixtures," *Journal of the Royal Statistical Society B*, 25, 235–260 (with discussion).

Smucker, B. J., del Castillo, E., and Rosenberger, J. L. (2012), "Model-Robust Two-Level Designs Using Coordinate Exchange Algorithms and a Maximin Criterion," *Technometrics*, 54, 367–375.

Van Nostrand, R. C. (1995), "Design of Experiments Where the Order of Addition is Important, ASA Proceedings, Physical and Engineering Section, 155–160. Alexandria, VA: American Statistical Association.

Voelkel, J. G. (2019), "The Design of Order-of-Addition Experiments," *Journal of Quality Technology*, 51, 230–241.

Winkler, P., Chen, J., and Lin, D. K. J. (2020), "The Construction of Optimal Design for Order-of-Addition Experiment via Threshold Accepting," in J. Fan, and J. Pan (eds), *Multivariate Analysis and Data Mining*, *Chapter 6*, New York: Springer, pp 93–109.

Yang, J.-F., Sun, F., and Xu, H. (2020), "A Component-Position Model, Analysis and Design for Order-of-Addition Experiments," *Technometrics*, , – .

Zhao, Y., Li, Z., and Zhao, S. (2020), "A New Method of Finding Component Orthogonal Arrays for Order-of-Addition Experiments," *Metrika*, , – .

Table 1. Design and data for an anti-tumor drug combination experiment (reproduced from Yang et al. 2020, Table 1, omitting the seventh run)

| Run | Permutation of drugs | | | Response |
|---|---|---|---|---|
| 1 | A | B | C | 26.7 |
| 2 | A | C | B | 35.3 |
| 3 | B | A | C | 32.4 |
| 4 | B | C | A | 48.7 |
| 5 | C | A | B | 35.9 |
| 6 | C | B | A | 37.6 |

Table 2. Summary results for data of Table 3 in Yang et al. (2020)

| Statistic | Model[§] | | | | |
|---|---|---|---|---|---|
| | PWO | tPWO | CP | RS | NN |
| Error d.f. | 16 | 16 | 13 | 14 | 11 |
| RMSE | 3.43 | 3.32 | 3.65 | 3.25 | 3.68 |
| AIC | 135.6 | 134.0 | 139.6 | 133.8 | 139.9 |
| BIC | 146.2 | 144.6 | 153.7 | 146.8 | 186.6 |
| $w_k$ | 0.171 | 0.376 | 0.023 | 0.410 | 0.020 |
| $avd(\sigma^2=1)$ | 0.52174 | 0.52174 | 0.78261 | 0.69565 | 0.95652 |
| $avd$ | 6.14128 | 5.75126 | 12.7599 | 7.36493 | 12.9380 |

[§] PWO = pairwise ordering; tPWO = tapered PWO using $z(h)=1/h$; CP = component position; RS = response surface; NN = nearest neighbour; the model comprises a dummy for the two component orthogonal arrays of the design.

Table 3. Predictions of $\eta_f$ for the best ten orders of application based on the model-average for the data of Table 3 in Yang et al. (2020)

| Permutation of components | | | | Model[§] | | | | | | | | | | MA[$] | | |
|---|---|---|---|---|---|---|---|---|---|---|---|---|---|---|---|---|
| | | | | PWO | | tPWO | | CP | | RS | | NN | | | | |
| | | | | Estimate | Rank | Estimate | Rank | Estimate | Rank | Estimate | Rank | Estimate | Rank | Estimate | Rank | s.e. |
| 1 | 3 | 4 | 2 | 54.85 | 2 | 54.71 | 2 | 52.97 | 5 | 55.60 | 1 | 55.17 | 1 | 55.07 | 1 | 1.977 |
| 3 | 1 | 4 | 2 | 54.76 | 3 | 54.57 | 3 | 53.70 | 2 | 54.72 | 2 | 53.26 | 3 | 54.62 | 2 | 1.926 |
| 3 | 4 | 1 | 2 | 55.57 | 1 | 56.10 | 1 | 53.87 | 1 | 50.89 | 7 | 53.89 | 2 | 53.78 | 3 | 3.089 |
| 3 | 1 | 2 | 4 | 52.38 | 6 | 52.35 | 5 | 53.34 | 3 | 53.91 | 3 | 50.73 | 7 | 52.99 | 4 | 2.072 |
| 1 | 3 | 2 | 4 | 52.47 | 5 | 51.98 | 6 | 52.60 | 6 | 53.08 | 4 | 52.45 | 5 | 52.54 | 5 | 1.975 |
| 3 | 4 | 2 | 1 | 53.15 | 4 | 52.95 | 4 | 53.17 | 4 | 50.20 | 9 | 52.16 | 6 | 51.84 | 6 | 2.354 |
| 3 | 2 | 4 | 1 | 50.77 | 7 | 50.91 | 7 | 49.39 | 9 | 50.96 | 6 | 52.83 | 4 | 50.91 | 7 | 1.938 |
| 3 | 2 | 1 | 4 | 49.96 | 8 | 50.67 | 8 | 49.73 | 8 | 50.85 | 8 | 50.32 | 8 | 50.59 | 8 | 1.943 |
| 1 | 4 | 3 | 2 | 46.90 | 11 | 47.08 | 11 | 50.93 | 7 | 51.01 | 5 | 49.39 | 10 | 48.79 | 9 | 2.737 |
| 4 | 1 | 3 | 2 | 47.71 | 9 | 48.34 | 9 | 45.67 | 13 | 46.20 | 11 | 49.66 | 9 | 47.32 | 10 | 2.173 |

§ PWO = pairwise ordering; tPWO = tapered PWO using $z(h) = 1/h$; CP = component position; RS = response surface; NN = nearest neighbour;
$ Model-averaged prediction using Akaike weights (see eg. 14); the model comprises a dummy for the two component orthogonal arrays of the design.

Table 4. Summary results for data of Table 4 in Yang et al. (2020)

| Statistic | Model[§] | | | | |
|---|---|---|---|---|---|
| | PWO[&] | tPWO | CP | RS | NN |
| Error d.f. | 28 | 28 | 22 | 25 | 19 |
| RMSE | 4.92 | 4.96 | 4.65 | 3.60 | 5.06 |
| AIC | 252.7 | 253.3 | 250.6 | 229.3 | 257.4 |
| BIC | 274.6 | 275.3 | 282.6 | 256.3 | 294.5 |
| $w_k$ | $8.27\times10^{-6}$ | $6.01\times10^{-6}$ | $2.38\times10^{-5}$ | 0.99996 | $7.90\times10^{-7}$ |
| $avd(\sigma^2=1)$ | 0.54250 | 0.55666 | 0.80672 | 0.66428 | 0.59773 |
| $avd$ | 13.1200 | 13.6788 | 12.9271 | 8.62647 | 15.2759 |

§ PWO = pairwise ordering; tPWO = tapered PWO using $z(h)=1/h$; CP = component position; RS = response surface; NN = nearest neighbour; the fitted model comprised a dummy for the two batches of the design.

Table 5. Predictions of $\eta_f$ for the best ten orders of application based on the model-average for the data of Table 4 in Yang et al. (2020)

| Permutation of components | | | | | Model[§] | | | | | | | | | | MA[$] | | |
|---|---|---|---|---|---|---|---|---|---|---|---|---|---|---|---|---|---|
| | | | | | PWO | | tPWO | | CP | | RS | | NN | | | | |
| | | | | | Estimate | Rank | Estimate | Rank | Estimate | Rank | Estimate | Rank | Estimate | Rank | Estimate | Rank | s.e. |
| 3 | 1 | 4 | 5 | 2 | 23.42 | 39 | 23.86 | 35 | 23.78 | 37 | 29.88 | 1 | 34.66 | 2 | 29.88 | 1 | 2.181 |
| 3 | 1 | 5 | 4 | 2 | 25.61 | 22 | 26.38 | 20 | 26.51 | 17 | 29.45 | 2 | 30.80 | 9 | 29.45 | 2 | 2.230 |
| 3 | 1 | 5 | 2 | 4 | 27.75 | 12 | 28.79 | 7 | 29.77 | 7 | 29.01 | 3 | 31.56 | 7 | 29.01 | 3 | 2.181 |
| 3 | 5 | 2 | 1 | 4 | 30.13 | 3 | 30.32 | 2 | 34.33 | 1 | 28.80 | 4 | 29.06 | 15 | 28.80 | 4 | 2.181 |
| 3 | 1 | 4 | 2 | 5 | 20.67 | 64 | 19.90 | 69 | 23.46 | 40 | 28.57 | 5 | 25.67 | 29 | 28.57 | 5 | 2.112 |
| 3 | 5 | 2 | 4 | 1 | 29.27 | 6 | 29.45 | 6 | 31.15 | 3 | 28.38 | 6 | 30.92 | 8 | 28.38 | 6 | 2.112 |
| 5 | 2 | 4 | 1 | 3 | 21.42 | 57 | 22.19 | 52 | 22.03 | 54 | 28.37 | 7 | 26.36 | 28 | 28.37 | 7 | 2.112 |
| 3 | 5 | 1 | 2 | 4 | 30.08 | 4 | 29.50 | 5 | 33.02 | 2 | 28.28 | 8 | 21.80 | 56 | 28.28 | 8 | 2.230 |
| 3 | 2 | 5 | 1 | 4 | 27.38 | 13 | 27.59 | 10 | 28.46 | 10 | 28.12 | 9 | 28.21 | 21 | 28.12 | 9 | 2.081 |
| 3 | 1 | 2 | 5 | 4 | 25.00 | 27 | 25.61 | 26 | 27.44 | 14 | 27.98 | 10 | 33.22 | 4 | 27.98 | 10 | 2.081 |

§ PWO = pairwise ordering; tPWO = tapered PWO using $z(h)=1/h$; CP = component position; RS = response surface; NN = nearest neighbour;

$ Model-averaged prediction using Akaike weights (see eg. 14); the fitted model comprised a dummy for the two batches of the design.

**Table A1**: Summary results for third-order RS models fitted to data of Table 3 in Yang et al. (2020)

| Statistic | Model | |
|---|---|---|
| | RS-3 special[§] | RS-3[$] |
| Error d.f. | 11 | 8 |
| RMSE | 3.42 | 3.58 |
| AIC | 136.4 | 136.9 |
| BIC | 152.9 | 157.0 |

[§] RS-3 special = special third-order response surface model (A7)

[$] RS-3 = third-order response surface model (A6)

**Table A2**: Summary results for data of Table 4 in Yang et al. (2020)

| Statistic | Model | |
|---|---|---|
| | RS-3 special[§] | RS-3[$] |
| Error d.f. | 17 | 11 |
| RMSE | 4.00 | 4.53 |
| AIC | 238.2 | 242.7 |
| BIC | 278.7 | 293.4 |

[§] RS-3 special = special third-order response surface model (A7)

[$] RS-3 = third-order response surface model (A6)

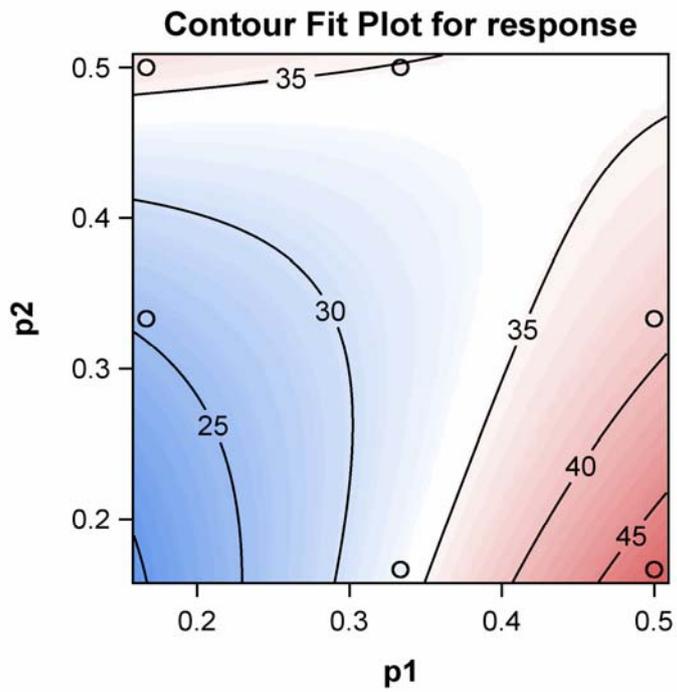

Figure 1. Contour plot for fit of model (10) plotted in the ($p_1$, $p_2$)-plane for data in Table 1.